\newcommand{\be}{\begin{equation}}
\newcommand{\ee}{\end{equation}}
\newcommand{\bea}{\begin{eqnarray}}
\newcommand{\eea}{\end{eqnarray}}

\newcommand{\itDelta}{{\it \Delta}}

\newcommand{\itPi}{{\it \Pi}}
\newcommand{\itSigma}{{\it \Sigma}}

\newcommand{\bra}{\langle}
\newcommand{\ket}{\rangle}
\newcommand{\dbra}{\bra \! \bra}
\newcommand{\dket}{\ket \! \ket}

\newcommand{\me}{\mbox{e}}

\newcommand{\bq}{{\bar q}}
\newcommand{\rK}{{\rm K}}

\newcommand{\rRe}{{\rm Re}}

\newcommand{\rN}{{\rm N}}
\newcommand{\rS}{{\rm S}}
\newcommand{\rNS}{{\rm NS}}

\documentclass{elsart}
\usepackage{epsfig}

\usepackage{amssymb}

\begin{document}

\begin{frontmatter}



\title{PDF of Velocity Fluctuation in 
Turbulence by a Statistics based on Generalized Entropy}



\author[label2]{Toshihico~Arimitsu}
and
\author[label3]{Naoko~Arimitsu}

\address[label2]{Institute of Physics, University of Tsukuba, Ibaraki 305-8571,
Japan}
\address[label3]{Graduate School of EIS, Yokohama Nat'l.~University, 
Yokohama 240-8501, Japan}

\begin{abstract}
An analytical formula for the probability density function (PDF) of 
the velocity fluctuation in fully-developed turbulence is derived, 
non-perturbatively, by assuming that its underlying statistics is 
the one based on the generalized measures of entropy, the R\'{e}nyi entropy 
or the Tsallis entropy. 
The parameters appearing in the PDF, including the index $q$ 
in the generalized measures, are 
determined self-consistently with the help of the observed 
value $\mu$ of the intermittency exponent.
The derived PDF explains quite well the experimentally observed density functions.
\end{abstract}

\end{frontmatter}


In this paper, we analyze the probability density function (PDF)
of the velocity fluctuations in turbulence measured in the experiments conducted by 
Lewis and Swinney~\cite{Lewis-Swinney99} and by Gotoh et al.~\cite{Gotoh01}
in terms of a generalized statistics constructed on
R\'{e}nyi's \cite{Renyi} or on Tsallis' entropy \cite{Tsallis88,Tsallis99}.
The R\'{e}nyi entropy has an information theoretical basis, and has 
the extensive characteristics as the usual thermodynamical entropy does.
On the other hand, the Tsallis entropy is non-extensive, and therefore
it provides us with one of the attractive trials for generalized statistical mechanics 
which deals with, for example, systems with long range correlations.
Note that the distribution functions which give an extremum of each entropy
have a common structure (see (\ref{Tsallis prob density}) for the present system)
in spite of different characteristics of these entropies.
The investigation of turbulence based on the generalized statistics was
set about by the present authors~\cite{AA} with an investigation 
of the p-model~\cite{Meneveau87a,Meneveau87b}.
Then, the treatment was sublimated by a self-consistent analysis following the usual 
procedure of ensembles in statistical mechanics, where it was shown that the derived 
analytical expression for the scaling exponents of the velocity structure function 
explains quite well the experimentally observed exponents~\cite{AA1,AA2,AA3,AA4}. 
It has been revealed also that, within the same line of 
the statistical mechanical treatment, the PDF
of the velocity fluctuations can be derived analytically~\cite{AA4,AA5}.
An input for the ensemble theoretical method is the experimentally measured
intermittency exponent. The parameters appearing in the ensemble theory
are determined self-consistently with the help of the physically meaningful
equations, i.e., the conservation of energy, the definition of the intermittency exponent
and the scaling relation relevant to the analysis by statistics based on
the generalized entropy~\cite{AA1,AA2,AA3,AA4,AA5}.
The present ensemble theoretical method rests on the following aspects:
1) the fully developed turbulence is the situation where the kinematic
viscosity can be neglected compared with the effect of the turbulent viscosity
that is a cause of intermittency; 2) in this situation, one can find an invariance
of the Navier-Stokes equation in certain scale transformation which contains
an arbitrary exponent that measures the degrees of singularity in velocity gradient;
3) for the distribution of the exponent, the statistics based on 
the generalized entropies is adopted, 
which was shown~\cite{AA2,AA3,AA4,AA5} to be a generalization of the statistics in
the log-normal model~\cite{Oboukhov62,K62,Yaglom}.
The statistical mechanical procedure based on the Boltzmann-Gibbs entropy
gives the results of the log-normal model, whereas the one based on
the generalized entropies provides us with the above mentioned 
new results~\cite{AA4}.



The basic equation describing fully developed turbulence is 
the Navier-Stokes equation
$
\partial {\vec u}/\partial t
+ ( {\vec u}\cdot {\vec \nabla} ) {\vec u} 
= - {\vec \nabla} \left(p/\rho \right)
+ \nu \nabla^2 {\vec u}
\label{N-S eq}
$
of an incompressible fluid, where $\rho$, $p$ and 
$\nu$ represent, respectively, the mass density, 
the pressure and the kinematic viscosity.
Our main interest is the correlation of 
measured time series of the streamwise velocity component $u$, 
say $x$-component of the fluid velocity field ${\vec u}$.
Within Taylor's frozen flow hypothesis, 
the quantity of our interest reduces to
$
\delta u(r) = \vert u(x+r) - u(x) \vert
$
representing the spatial difference of the component $u$.
Then, the Reynolds number $\rRe$ of the system is given by 
$
{\rm Re} = \delta u_0 \ell_0/\nu = ( \ell_0/\ell_{\rK} )^{4/3}.
$
Here, we used 
$
\delta u_0=\left(\epsilon \ell_0\right)^{1/3}
$
with the Kolmogorov scale~\cite{K41}: 
$
\ell_{\rK} = ( \nu^3/\epsilon )^{1/4} ( = \eta )
$,
and the energy input rate $\epsilon$ to the largest eddies
of size $\ell_0$.
Here, we introduced the notation
$
\delta u_n = \delta u(\ell_n)
$
representing the velocity difference across a distance $r \sim \ell_n$.

For high Reynolds number limit
$
\rRe \gg 1
$, 
or for the situation where the effect of the term containing 
the kinematic viscosity in the Navier-Stokes equation can be effectively neglected,
we see that the equation has an invariance 
under the scale transformation~\cite{Meneveau87b}:
${\vec r} \rightarrow {\vec r}\ '$, ${\vec u} \rightarrow {\vec u}\ '$, 
$t \rightarrow t'$ and $\left(p/\rho\right) \rightarrow \left(p/\rho\right)'$
with
$
{\vec r}\ ' = \lambda {\vec r}
$,
$
{\vec u}\ ' = \lambda^{\alpha/3} {\vec u}
$,
$
t' = \lambda^{1- \alpha/3} t
$,
$
\left(p/\rho\right)' = \lambda^{2\alpha/3}
\left(p/\rho\right)
$.
Here, $\alpha$ is an arbitrary real quantity.
Then, we have 
\be
\delta u_n / \delta u_0 = (\ell_n / \ell_0)^{\alpha/3},\qquad
\epsilon_n/\epsilon = \left(\ell_n/\ell_0 \right)^{\alpha -1}
\label{u/epsilon-alpha}
\ee
where $\epsilon_n$ represents the rate of transfer of energy per unit mass 
from eddies of size $\ell_n$ to eddies of size $\ell_{n+1}$.
We have put $\epsilon_0 = \epsilon$.
The first equation in (\ref{u/epsilon-alpha}) leads to a singularity in 
the velocity gradient~\cite{Benzi84}
$
\left\vert \partial u(x)/\partial x \right\vert 
= \lim_{\ell_n \rightarrow 0} \vert u(x+\ell_n) - u(x) \vert/\ell_n
= \lim_{\ell_n \rightarrow 0} \delta u_n/\ell_n
$
for $\alpha < 3$, since 
$
\delta u_n/\ell_n^{\alpha/3} = {\rm constant}
$.
The degrees of the singularity are specified by the exponent $\alpha$.
The singularities in velocity gradient fill the physical space of dimension $d$ 
with a fractal dimension $F(\alpha)$. 
Then, the probability $P(\alpha) d\alpha$ to find a point to be 
singular is specified by the distribution function of singularities 
$
P(\alpha) \propto \delta_n^{d-F(\alpha)}
\label{P-fd}
$,
which leads through the second relation in (\ref{u/epsilon-alpha}) that 
the energy transfer rate $\epsilon_n$ is a fluctuating quantity in space. 
The conservation of energy in the energy-transfer cascade requires that
\be
\left\bra \epsilon_n \right\ket = \epsilon
\label{cons of energy}
\ee
where $\bra \cdots \ket$ is defined by
$
\bra \cdots \ket = \int d\alpha \cdots P(\alpha)
$.
The intermittency observed in the systems of fully-developed turbulence 
is attributed to the distribution of the singularities 
in each different size 
$
\ell_n= \ell_0 \delta_n
$ 
$(n = 0, 1, 2,\cdots )$
of constituting eddies.
Here, $\ell_n$ represents the diameter of the $n$th eddy, and
is determined by the number of steps $n$ of the cascade~\cite{Frisch78}.
It is assumed by putting 
$\delta_n = \delta^{-n}$
$
(\delta = 2)
$
that, at each step of the cascade, eddies break into $\delta$ pieces with 
$1/\delta$ diameter. 
The definition of the intermittency exponent $\mu$ is
given by
\be
\bra \epsilon_n^2 \ket 
= \epsilon^2 \delta_n^{-\mu}.
\label{def of mu}
\ee


It is convenient to introduce multifractal spectrum $f(\alpha)$ by
$
f(\alpha) = F(\alpha) - (d-1)
\label{f}
$
which gives us
$
P(\alpha) \propto \delta_n^{1-f(\alpha)}
\label{P-f}
$.


Regarding an eddy as a cluster having the same size 
as the eddy \cite{AA,AA4},
we can derive the {\it scaling relation} \cite{AA1,AA2,AA3,AA4} 
\be
1/(1-q) = 1/\alpha_- - 1/\alpha_+,
\label{scaling relation}
\ee
where $\alpha_{\pm}$ are defined by $f(\alpha_{\pm})=0$.
This scaling relation is a manifestation of the mixing property.
Note that (\ref{scaling relation}) is a generalization of 
the scaling relation derived first in~\cite{Lyra98,Costa} to
the case where the multifractal spectrum $f(\alpha)$ has
negative values~\cite{Chhabra91}.



Let us find out an appropriate distribution function 
$
P^{(n)}(\alpha) \propto [ P^{(1)}(\alpha) ]^n
$
of the degrees of singularity $\alpha$ at the $n$th step of the cascade.
Here, we are assuming the independence of each step.
In order to discover $P^{(1)}(\alpha)$, we will
take an extremum of the measure of an appropriate generalized entropy 
with the two constraints, i.e., the conservation of probability:
$
\int d\alpha P(\alpha) = \mbox{constant}
\label{cons of prob}
$,
and the fixing of the $q$-variance:
$
\sigma_q^2 = \bra (\alpha- \alpha_0)^2 \ket_{q}
= (\int d\alpha P(\alpha)^{q} 
(\alpha- \alpha_0 )^2 ) / \int d\alpha P(\alpha)^{q}
\label{q-variance}
$.
By adopting the measure of the R\'{e}nyi entropy~\cite{Renyi} or of
the Tsallis-Havrda-Charvat (THC) entropy~\cite{Tsallis88,Havrda-Charvat} 
as an appropriate one, we have the distribution function
of singularities in the form~\cite{AA1,AA2,AA3,AA4,AA5}
\be
P^{(n)}(\alpha)\ d\alpha = \left( Z^{(n)} \right)^{-1}
\left[ 1 - (\alpha - \alpha_0)^2 / (\itDelta \alpha )^2 
\right]^{n/(1-q)} d\alpha 
\label{Tsallis prob density}
\ee
with 
$
\itDelta \alpha = \left(2X / \left[(1-q) \ln 2 \right] \right)^{1/2}
$
and an appropriate partition function
$
Z^{(n)}
$.
Note that since the R\'{e}nyi entropy and the THC entropy reduce to 
the Boltzmann-Gibbs entropy in the limit
$
q \rightarrow 1
$,
the distribution function (\ref{Tsallis prob density}) 
reduces to the Boltzmann-Gibbs distribution function in that limit.
The multifractal spectrum corresponding to 
the distribution function of singularities is given by~\cite{AA1,AA2,AA3,AA4,AA5}
\be
f(\alpha) = 1 + (1-q)^{-1} \log_2 \left[ 1 - \left(\alpha - \alpha_0\right)^2
\Big/ \left(\Delta \alpha \right)^2 \right]
\label{Tsallis f-alpha}
\ee
where we put $f(\alpha_0) = 1$, since the fractal dimension of 
the multifractal sets is one for the present experimental setup of the system.
Note that the multifractal spectrum does not depend on 
the number $n$ of steps in the cascade, which is a manifestation 
of the assumption that each step is independent.


The mass exponent $\tau(\bq)$ is defined by
$
\delta_n^{-d} \bra (\epsilon_n \ell_n^d )^{\bq} \ket
= (\epsilon \ell_0^d )^{\bq} \delta_n^{-\tau(\bq) +(\bq -1)(d-1)}
\label{def of mass exponent}
$
where $\epsilon_n \ell_n^d$ is the total dissipation in volume $\ell_n^d$, and
the factor $\delta_n^{-d}$ is the number of boxes of volume $\ell_n^d$
in the region of volume $\ell_0^d$.\footnote{This definition is
a bit different from the one in \cite{AA,AA1} for general $d$.
}
Substituting
$
P(\alpha) \propto \delta_n^{1-f(\alpha)}
$
into the definition, we have
$
\delta_n^{-\tau(\bq)}
= \delta_n^{-d}
\int_{\alpha_{\rm min}}^{\alpha_{\rm max}} 
d\alpha' P(\alpha')\ \delta_n^{\bq (\alpha' -1)}
\propto \delta_n^{\bq \alpha_{\bq} - f(\alpha_{\bq})}
\label{tau-f derivation}
$
by estimating the integral in the limit $\delta_n \ll 1$, 
where $\alpha_{\bq}$ is obtained by solving
\be
\bq = df(\alpha)/d\alpha.
\label{bq-alpha}
\ee
Then, comparing the exponents of $\delta_n$ in the estimate of the integral, we know
that the relation between the multifractal spectrum $f(\alpha)$ and the mass exponent 
$\tau(\bq)$ is given by
\be
f(\alpha) = \bq \alpha + \tau(\bq)
\label{f-tau}
\ee
with $\alpha$ being determined by (\ref{bq-alpha}).
From (\ref{bq-alpha}) and (\ref{f-tau}), we can derive
\be
\alpha = - d\tau(\bq)/d\bq.
\label{alpha-bq}
\ee
Note that  (\ref{bq-alpha}), (\ref{f-tau}) and (\ref{alpha-bq})
constitute a kind of Legendre transformation between 
the mass exponent $\tau(\bq)$ and the multifractal spectrum
$f(\alpha)$.
Equation (\ref{bq-alpha}) is solved, analytically, to give us
$
\alpha_{\bq} = \alpha_0 - 2\bq X /(1 + \sqrt{C_{\bq}} )
\label{alpha-qbar}
$
with
$
{C}_{\bq}= 1 + 2 \bq^2 (1-q) X \ln 2
\label{cal D}
$.
Here, we chose the solution which is finite for $q \rightarrow 1$.
Then, $\alpha_{\rm max}=\alpha(\bq = -\infty)$
and $\alpha_{\rm min}=\alpha(\bq = +\infty)$ are given by
$
\alpha_{\rm max} - \alpha_0 
= \alpha_0 - \alpha_{\rm min}
= \itDelta \alpha
$.
We now know that 
$
\sigma_q^2 = (\itDelta \alpha )^2 (1-q)/(3-q)
$.
The conditions (\ref{cons of energy}) and (\ref{def of mu}) can be 
written, respectively, in terms of the mass exponent as
$
\tau(1) = 0
\label{cons of energy by tau}
$
and
$
\mu = 1 + \tau(2)
\label{def of mu by tau}
$.


The three quantities $\alpha_0$, $X$ and $q$ are determined by
the three independent equations
(\ref{cons of energy}), (\ref{def of mu}) 
and (\ref{scaling relation}), once the value of 
the intermittency exponent $\mu$ is given.
The scaling relation (\ref{scaling relation}) can be solved, 
by substituting
$
\alpha_{\pm} = \alpha_0 \pm \sqrt{2bX} 
\label{alpha mp}
$
satisfying
$
f_{\rm T}(\alpha_{\pm}) = 0,
$
in the form
$
\sqrt{2X} =[ (\alpha_0^2 +(1-q)^2)^{1/2}
- (1-q) ] / \sqrt{b},
\label{X}
$
or
$
\alpha_0= [2bX + 2(1-q) \sqrt{2bX}]^{1/2},
$
with
$
b=(1-2^{-(1-q)})/[(1-q)\ln2]
$.



Let us now derive the PDF for the velocity fluctuation
$\itPi^{(n)}(x_n)$ with $-\infty < x_n < \infty$, where
$
\vert x_n \vert = \delta u_n / \delta u_0.
$
It may be reasonable to assume that the probability $\itPi^{(n)}(x_n) dx_n$
to find the scaled velocity fluctuation $x_n$ in the range 
$x_n \sim x_n+dx_n$ can be divided into two parts~\cite{AA5}:
\be
\itPi^{(n)}(x_n) dx_n = \itPi_{\rN}^{(n)}(x_n) dx_n 
+ \itPi_{\rS}^{(n)}(\vert x_n \vert) dx_n.
\ee
The first term $\itPi_{\rN}^{(n)}(x_n)$, assumed to come from thermal or 
dissipative fluctuation caused by the kinematic viscosity, will be considered 
later in this paper after estimating the moments of the velocity fluctuation.
The second term $\itPi_{\rS}^{(n)}(\vert x_n \vert)$ stems
from the multifractal distribution of singularities in the
velocity gradient given in (\ref{Tsallis prob density}), i.e., 
$
\itPi_{\rS}^{(n)}(\vert x_n \vert) dx_n = P^{(n)}(\alpha) d \alpha.
$
This contribution may be related to the fluctuations caused by
the turbulent viscosity.
The condition
$
\int_{-\infty}^{\infty} d x_n \itPi^{(n)}(x_n) = 1
\label{normalization of PDF}
$
determines the normalization factor $Z^{(n)}$ here.
For this singular part, the relation between $\vert x_n \vert$ and $\alpha$ is given by
(\ref{u/epsilon-alpha}), i.e., 
$
\vert x_n \vert = \delta_n^{\alpha/3}
$,
and, therefore, we have an explicit expression of 
$\itPi_{\rS}^{(n)}(\vert x_n \vert)$ in the form~\cite{AA4,AA5}
\be
\itPi_{\rS}^{(n)}(\vert x_n \vert) dx_n 
= \left[3\Big/\left(Z^{(n)} \vert \ln \delta_n \vert 
\right) \right] G^{(n)}(\vert x_n \vert) dx_n
\label{PiS}
\ee
with
$
G^{(n)}(x) = x^{-1} [1 - (\alpha(x) -\alpha_0)^2 /(\Delta \alpha)^2]^{n/(1-q)}
$
and
$
\alpha(x) = 3\ln x /\ln \delta_n
$.
Note that the singular part of the PDF has a dependence
on $\ln \delta u_n$.


The $m$th moment can be calculated as~\cite{AA5}
$
\dbra \vert x_n \vert^m \dket = \int_{-\infty}^{\infty} dx_n\  
\vert x_n \vert^m \itPi^{(n)}(x_n)
= 2 \gamma_m^{(n)} + \left(1-2\gamma_0^{(n)} \right)
a_m \ \delta_n^{\zeta_m}
$
with
$
2\gamma_m^{(n)} = \int_{-\infty}^{\infty} dx_n\ 
\vert x_n \vert^m \itPi_\rN^{(n)}(x_n)
$
and
$
a_{3\bq} = ( 2 / [\sqrt{C_{\bq}} ( 1+ \sqrt{C_{\bq}} ) ] )^{1/2}
$.
Here, 
$
\zeta_m = \alpha_0 m/3 - 2Xm^2/[9 (1+\sqrt{{C}_{m/3}} )] 
- [1-\log_2 (1+\sqrt{{C}_{m/3}} ) ]/(1-q)
\label{zeta}
$
is the so-called scaling exponents of velocity structure function,
whose expression was derived first in \cite{AA1,AA2,AA3,AA4,AA5}.

\begin{figure}[thbp]
\begin{center}
\leavevmode
\epsfxsize=70mm
\epsfbox{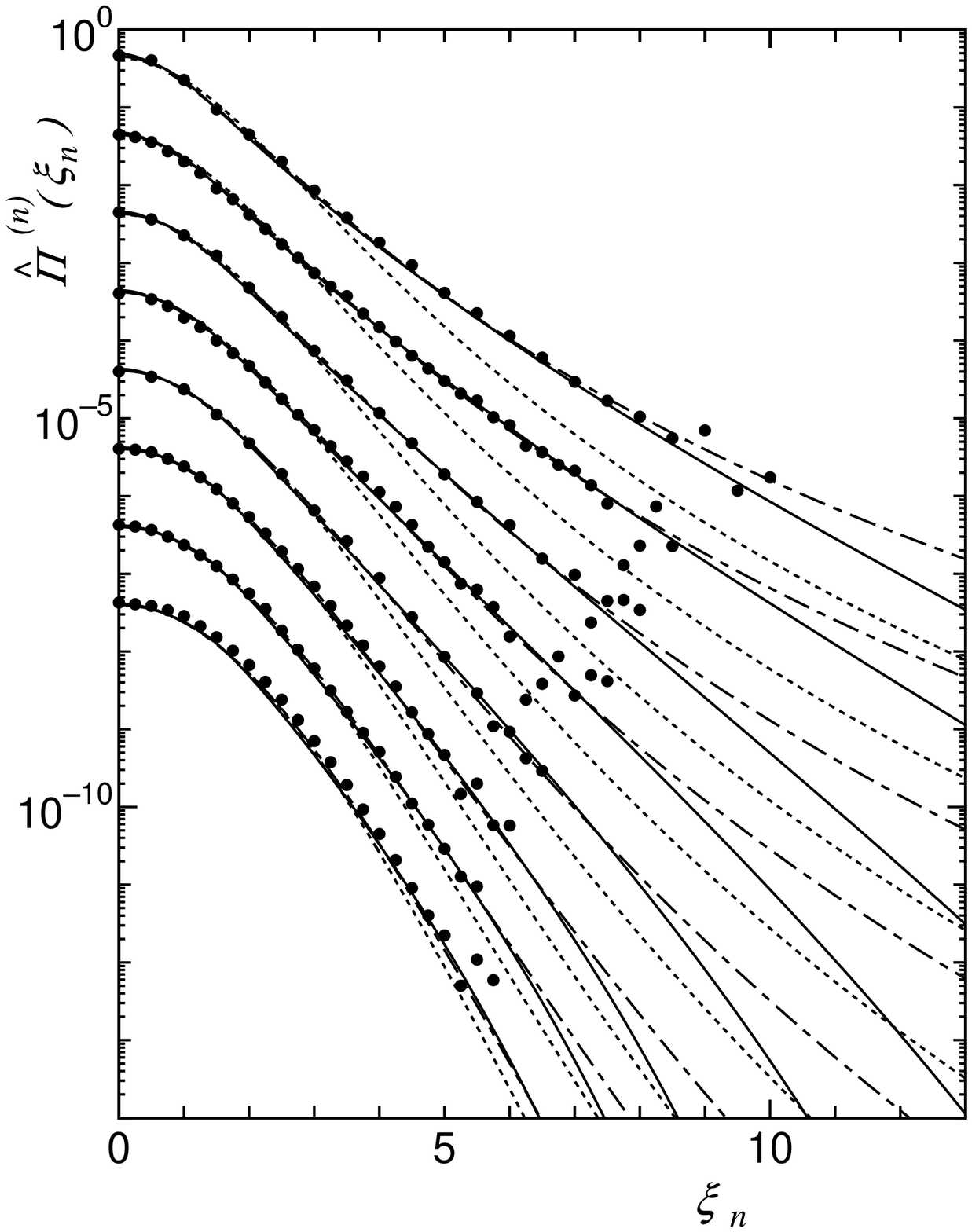}
\caption{Experimentally measured PDF of the velocity fluctuations by 
Lewis and Swinney \cite{Lewis-Swinney99} for $R_\lambda =262$ ($\rRe = 540\ 000$) 
and $\mu = 0.28$
are compared with the present theoretical results $\hat{\itPi}^{(n)}(\xi_n)$
given by (\ref{Pi-xi theoretical large}) and (\ref{Pi-xi theoretical small}).
Closed circles are the symmetrized points obtained by taking 
averages of the left and the right hand sides of measured PDF. 
For the experimental data, the distances $r/\eta = \ell_n/\eta$ 
are, from top to bottom: 11.6, 23.1, 46.2, 92.5 208, 399, 830 and 1440. 
Solid lines represent the curves given by the present theory with 
the self-consistently determined parameters $\alpha_0 = 1.162$, $X = 0.334$
and $q = 0.471$.
For the present theoretical curves (\ref{Pi-xi theoretical large}) and 
(\ref{Pi-xi theoretical small}), the number of steps in the cascade $n$ are,
from top to bottom: 14, 13, 11, 10, 9.0, 8.0, 7.5 and 7.0.
Dotted lines are the curves given by directly adjusting the PDF 
(\ref{Beck dist}) for $\theta = 1$, 
whereas dotted-dashed lines are the ones for $\theta = 2 - q'$.
For these adjusted PDF's, the indexes $q'$'s depend on the distances
$r/\eta$, and the adjusted values of $q'$ are, from top to bottom: 
1.168, 1.150, 1.124, 1.105, 1.084, 1.065, 1.055 and 1.038.
For better visibility, each PDF is shifted by -1 unit along the vertical axis.}
\label{Fig: Pi-xi ls}
\end{center}
\end{figure}

In the following, we will derive the symmetric part of the 
PDF of the velocity difference.
The origin of the asymmetry in the PDF
may be attributed to dissipative evolution of eddies or to 
experimental setup and situations.
Let us introduce new variable $\xi_n = x_n / \dbra x_n^2 \dket^{1/2}$ scaled 
by the variance of the velocity fluctuation, and introduce the probability density 
function $\hat{\itPi}_{\rS}^{(n)}(\xi_n)$ for the singular part 
with this new variable through the relation
$
\hat{\itPi}_{\rS}^{(n)}(\vert \xi_n \vert) d\xi_n 
= \itPi_{\rS}^{(n)}(\vert x_n \vert) dx_n
$.
It is a symmetric function satisfying 
$
\hat{\itPi}_{\rS}^{(n)}(\xi_n) = \hat{\itPi}_{\rS}^{(n)}(-\xi_n)
$.
The relation between the new variable and $\alpha$ is given by
$
\vert \xi_n \vert = \delta u_n / \bra \delta u_n^2 \ket^{1/2} 
= \bar{\xi}_n \delta_n^{\alpha /3 -\zeta_2 /2}
$
with
$
\bar{\xi}_n = [2 \gamma_2^{(n)} \delta_n^{-\zeta_2} + (1-2\gamma_0^{(n)} ) 
a_2 ]^{-1/2}
$
and the explicit form of $\hat{\itPi}_{\rS}^{(n)}(\vert \xi_n \vert)$ in terms of 
$\alpha$ reduces to
\be
\hat{\itPi}_{\rS}^{(n)}(\vert \xi_n \vert) = \bar{\itPi}_{\rS}^{(n)}
\delta_n^{\zeta_2/2 -\alpha/3 +1 -f(\alpha)}
\label{Pi-alpha}
\ee
with
$
\bar{\itPi}_{\rS}^{(n)} = 3 (1-2\gamma_0^{(n)} )
/ (2 \bar{\xi}_n \sqrt{2\pi X \vert \ln \delta_n \vert} ).
$
We see from (\ref{Pi-alpha}) that when the condition
$
\zeta_2/2 -\alpha/3 +1 -f(\alpha) = 0
\label{cond of alpha}
$
is satisfied, the strong $n$-dependence of $\hat{\itPi}_{\rS}^{(n)}(\xi_n)$ 
may disappear.
Corresponding to the two solutions $\alpha^*$ and $\alpha^\dagger$ satisfying 
$\alpha^* < \alpha^\dagger$ for this condition, we have
$
\xi_n^\dagger = \bar{\xi_n}\ \delta_n^{\alpha^\dagger/3 - \zeta_2/2}
$
and
$
\xi_n^* = \bar{\xi_n}\ \delta_n^{\alpha^*/3 - \zeta_2/2}
$
satisfying $0 < \xi^\dagger < \xi^*$.

\begin{figure}[thbp]
\begin{center}
\leavevmode
\epsfxsize=70mm
\epsfbox{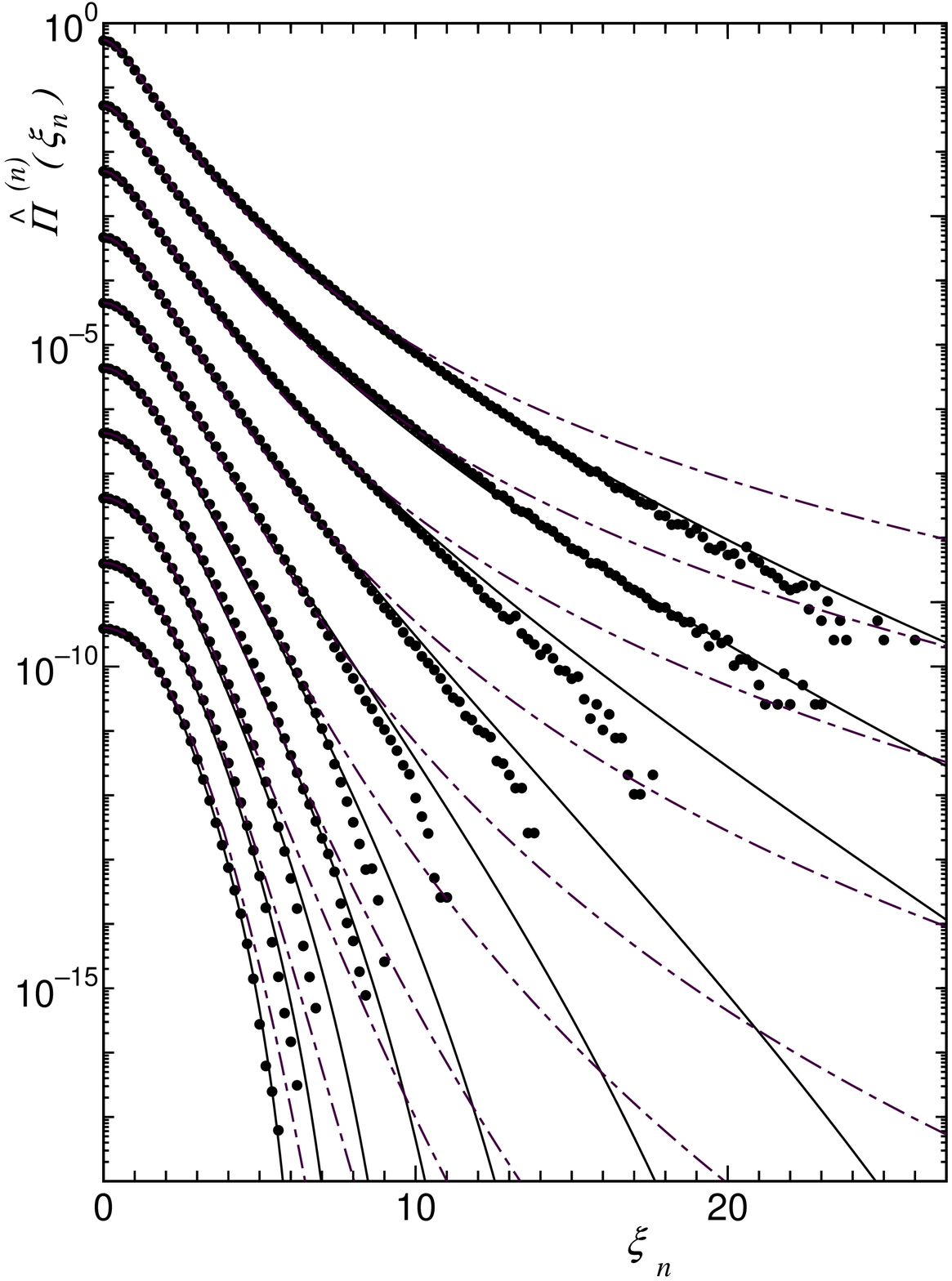}
\caption{Experimentally measured PDF of the velocity fluctuations by 
Gotoh et al.\ \cite{Gotoh01} for $R_\lambda =381$ and $\mu = 0.25$
are compared with the present theoretical results $\hat{\itPi}^{(n)}(\xi_n)$
given by (\ref{Pi-xi theoretical large}) and (\ref{Pi-xi theoretical small}).
Closed circles are the symmetrized points obtained by taking 
averages of the left and the right hand sides of measured PDF. 
For the experimental data, the distances $r/\eta = \ell_n/\eta$ 
are, from top to bottom: 2.38, 4.76, 9.52, 19.0, 38.1, 76.2, 152, 305,
609 and 1220. 
Solid lines represent the curves given by the present theory with 
the self-consistently determined parameters $\alpha_0 = 1.445$, $X = 0.298$
and $q = 0.413$.
For the present theoretical curves (\ref{Pi-xi theoretical large}) and 
(\ref{Pi-xi theoretical small}), the number of steps in the cascade $n$ are,
from top to bottom: 21, 19, 17, 14, 12, 10, 9.0, 8.0, 7.0 and 6.0.
Dotted-dashed lines are the curves given by (\ref{Beck dist}) for $\theta = 2 - q'$.
For this directly adjusted PDF, the indexes $q'$'s depend on the distances
$r/\eta$, and the adjusted values of $q'$ are, from top to bottom: 
1.215, 1.195, 1.175, 1.145, 1.115, 1.105, 1.095, 1.060, 1.030 and 1.005.
For better visibility, each PDF is shifted by -1 unit along the vertical axis.}
\label{Fig: Pi-xi gotoh}
\end{center}
\end{figure}

By making use of the point $\xi_n^*$ having only weak $n$-dependence, we divide 
$\hat{\itPi}^{(n)}(\xi_n)$ into two parts:
$
\hat{\itPi}^{(n)}(\xi_n) = \hat{\itPi}_{<*}^{(n)}(\xi_n)
$
for $\vert \xi_n \vert \leq \xi_n^*$, and
$
\hat{\itPi}^{(n)}(\xi_n) = \hat{\itPi}_{*<}^{(n)}(\xi_n)
$
for $\xi_n^* \leq \vert \xi_n \vert$.
For smaller velocity fluctuation, we assume that the PDF has Gaussian form:
$
\hat{\itPi}_{<*}^{(n)}(\xi_n) = \hat{\itPi}_{\rN}^{(n)}(\xi_n) 
+ \hat{\itPi}_{\rS}^{(n)}(\vert \xi_n \vert)
= (Z_{\rNS}^{(n)} )^{-1} \me^{-\xi_n^2/2\itSigma_n^2}
$,
whereas, for larger velocity fluctuation, we assume that it is given by the distribution
function for singularities~\cite{AA5}:
\be
\hat{\itPi}_{*<}^{(n)}(\xi_n) = \hat{\itPi}_{\rS}^{(n)}(\vert \xi_n \vert).
\label{Pi-xi theoretical large}
\ee
Requiring $\hat{\itPi}_{<*}^{(n)}(\xi_n)$ and $\hat{\itPi}_{*<}^{(n)}(\xi_n)$ 
at $\xi_n^*$ to have the same value and the same 
derivative there, we can determine the variance $\itSigma_n^2$ and 
the normalization factor $Z_{\rNS}^{(n)}$, respectively, as
$
\itSigma_n^2 = (\xi^* )^2 / [ 1+ 3f'(\alpha^*) ]
$
and
$
(Z_{\rNS}^{(n)} )^{-1} = \bar{\itPi}_{\rS}^{(n)} 
\me^{[1+3f'(\alpha^*)]/2}
$.
Then, we see that $\hat{\itPi}_{<*}^{(n)}(\xi_n)$ becomes~\cite{AA5}
\be
\hat{\itPi}_{<*}^{(n)}(\xi_n) = \bar{\itPi}_{\rS}^{(n)} 
\exp\left\{-\left[1+3f'(\alpha^*)\right] \left[\left(\xi_n/\xi_n^* \right)^2 -1 \right]
\Big/2 \right\}.
\label{Pi-xi theoretical small}
\ee



Figs.~\ref{Fig: Pi-xi ls} and \ref{Fig: Pi-xi gotoh} are
the comparisons of the PDF $\hat{\itPi}^{(n)}(\xi_n)$,
(\ref{Pi-xi theoretical large}) and (\ref{Pi-xi theoretical small}),
obtained by the present theory with the experiments conducted by
Lewis and Swinney~\cite{Lewis-Swinney99} for $R_\lambda =262$ ($\rRe = 540\ 000$) 
and $\mu = 0.28$, and by Gotoh et al.~\cite{Gotoh01} 
for $R_\lambda =381$ and $\mu = 0.25$.
In both figures, solid lines represent the curves by the present theory, 
whereas closed circles are the experimental data 
symmetrized by taking averages of the left and the right hand sides of 
measured PDF's.
In Fig.~\ref{Fig: Pi-xi ls} (Fig.~\ref{Fig: Pi-xi gotoh}),
the distances $r/\eta = \ell_n/\eta$ are, from top to bottom: 
11.6, 23.1, 46.2, 92.5 208, 399, 830 and 1440 
(2.38, 4.76, 9.52, 19.0, 38.1, 76.2, 152, 305, 609 and 1220)
for the experimental data,
on the other hand, the number $n$ of steps in the cascade
are, from top to bottom:
14, 13, 11, 10, 9.0, 8.0, 7.5 and 7.0
(21, 19, 17, 14, 12, 10, 9.0, 8.0, 7.0 and 6.0)
for our theoretical lines. 
We observe an excellent agreement between the measured PDF and 
the analytical formula of PDF derived by the present self-consistent theory.

The present analysis is based on the assumptions that 
the distribution of the singularity indexes $\alpha$ is multifractal and
that its statistics is based on the generalized measures of entropy, i.e.,
the R\'{e}nyi entropy or the THC entropy.
Note that within the present analysis we cannot distinguish which of 
the measures of entropies is responsible for the system of turbulence.
The latter may be one of the exciting future problems both for
theoreticians and for experimentalists.
It is also one of the attractive future problems to find out 
some connection between the present ensemble approach and
the one based on dynamical aspect starting with the Navier-Stokes equation,
e.g., the dynamical renormalization group approach, 
the mode-mode coupling theoretical approach and so on.


Before closing the paper, we should mention something about 
the usual procedures to analyze experimentally measured PDF based on
generalized statistics, e.g., Tsallis statistics which provides us with one of 
the possibilities for an abstract statistical mechanics 
for non-Gibbsian statistics that are being observed 
in various fields (refer to \cite{Tsallis99} for more information on 
recent progress).
Usually, one is apt to write down Tsallis-type PDF directly 
without any fundamental consideration when he or she is going to analyze measured PDF.
Let us explain it in the context of the present analysis of turbulence.
Assuming, from the beginning, that the PDF of the velocity fluctuations is 
of Tsallis-type, one may write it in the form
\be
\hat{\itPi}_B(\xi_n) \propto \left\{ 1 - (1-q') \vert \xi_n 
\vert^{2\theta}/(5-3q') \right\}^{1/(1-q')}
\label{Beck dist}
\ee
for the measured PDF symmetrized in the sense given in this paper.
Here, $q'$ is the index appearing in the Tsallis or the R\'{e}nyi entropies
which may have different meaning from our $q$ used in the present paper,
and $\theta$ is an extra parameter determined in order to adjust 
(\ref{Beck dist}) to the measured PDF.
Note that the PDF (\ref{Beck dist}) of the velocity fluctuations itself
has a power-law decay for large $\delta u_n$, while the present PDF 
does not have simple power-law decay but
has a bit complicated decay containing $\ln \delta u_n$ as can be seen 
in the expression (\ref{PiS}).
In Fig.~\ref{Fig: Pi-xi ls}, we put (\ref{Beck dist}) with $\theta = 1$
by dotted lines for comparison which is essentially the same as done
by Beck in \cite{Beck00}. We see that this PDF cannot explain
the measured one.
Following the analysis done in \cite{Beck-Lewis-Swinney01},
we took $\theta = 2-q'$, and put (\ref{Beck dist})
in Figs.~\ref{Fig: Pi-xi ls} and \ref{Fig: Pi-xi gotoh} for this case by
dotted-dashed lines just for comparison.
As can be observed in Fig.~\ref{Fig: Pi-xi ls}, we cannot see big difference
between our present PDF, 
(\ref{Pi-xi theoretical large}) and (\ref{Pi-xi theoretical small}),
given by solid lines and the PDF, (\ref{Beck dist}) with $\theta = 2-q'$,
given by dotted-dashed lines,
since the accuracy of experimental data for PDF was only of order $10^{-5}$.
While in Fig.~\ref{Fig: Pi-xi gotoh} the PDF was measured 
up to the order of $10^{-9}$, and we clearly see the difficulty in analyzing
the data by the PDF given by Beck and others in \cite{Beck-Lewis-Swinney01}.
There may be a possibility to remedy it by adjusting the dependence of $\theta$
on $q'$ further, whose physical meaning, however, we do not know.
In addition to this, the analysis based on (\ref{Beck dist})
has another difficulty, i.e., the index $q'$ depends on the distance
$r/\eta$ whose dependence is listed in the captions 
in Figs.~\ref{Fig: Pi-xi ls} and \ref{Fig: Pi-xi gotoh}.
There might exist a possible explanation about the $r/\eta$ dependence
of $q'$ by relating it to the non-extensivity of Tsallis entropy \cite{Beck01preprint}.
In any case, the present investigation tells us that one should choose 
appropriate variables before applying generalized statistics. 
Note that the values of $q'$ satisfy $q' > 1$, in contrast to the values of
$q$ for the present analysis where they satisfy $q<1$.
It is worth mentioning here that 
the present ensemble theoretical approach~\cite{AA1,AA2,AA3,AA4,AA5} is quite general,
in that the expression of the multifractal spectrum (\ref{Tsallis f-alpha}) based on 
a statistical mechanical procedure may provide an example for 
the phenomenological "entropy function" $S(z)$ with $z = 1-\alpha$ in the problem of 
large deviation statistics~\cite{Fujisaka}. 
An analysis in this direction will be reported elsewhere in the near future.

The authors would like to thank Prof.~C.~Tsallis and 
Dr.~A.K.~Rajagopal for their fruitful comments with encouragement,
and Prof.~T.~Gotoh for his kindness to send them the data of 
his numerical simulations.


\begin{thebibliography}{00}




\bibitem{Lewis-Swinney99} G.S.~Lewis and H.L.~Swinney, Phys. Rev. E {\bf 59}, 5457 (1999).
\bibitem{Gotoh01} T.~Gotoh, D.~Fukayama and T.~Nakano (2001) preprint.
\bibitem{Renyi} A.~R\'{e}nyi, {\it Proc.\ 4th Berkeley Symp.\ 
		Maths.\ Stat.\ Prob.} {\bf 1}, 547 (1961).
\bibitem{Tsallis88} C.~Tsallis, J. Stat. Phys. {\bf 52}, 479 (1988).
\bibitem{Tsallis99} C.~Tsallis, Braz. J. Phys. {\bf 29}, 1 (1999).
	On the subject see also at http://tsallis.cat.cbpf.br/biblio.htm.
\bibitem{AA} T.~Arimitsu and N.~Arimitsu, Phys. Rev. E {\bf 61},
		 3237 (2000).
\bibitem{Meneveau87a} C.~Meneveau and K.~R.~Sreenivasan,
		Phys. Rev. Lett. {\bf 59}, 1424 (1987).
\bibitem{Meneveau87b} C.~Meneveau and K.~R.~Sreenivasan,
		Nucl. Phys. B (Proc. Suppl.) {\bf 2}, 49 (1987).
\bibitem{AA1} T.~Arimitsu and N.~Arimitsu, J. Phys. A: Math. Gen. {\bf 33}, 
		L235 (2000) [{\footnotesize CORRIGENDUM}: {\bf 34}, 673 (2001)].
\bibitem{AA2} T.~Arimitsu and N.~Arimitsu, Chaos, Solitons and Fractals 
		{\bf 13}, 479 (2002).
\bibitem{AA3} T.~Arimitsu and N.~Arimitsu, Prog.~Theor.~Phys. {\bf 105}, 355
		(2001).
\bibitem{AA4} T.~Arimitsu and N.~Arimitsu, Physica A {\bf 259}, 177 (2001).
\bibitem{AA5} N.~Arimitsu and T.~Arimitsu, cond-mat/0109132 (2001).
\bibitem{Oboukhov62} A.~M.~Oboukhov, J. Fluid Mech. {\bf 13},
		77 (1962).
\bibitem{K62} A.~N.~Kolmogorov, J. Fluid Mech. {\bf 13},
		82 (1962).
\bibitem{Yaglom} A.~M.~Yaglom, Sov. Phys. Dokl. {\bf 11},
		26 (1966).
\bibitem{K41} A.~N.~Kolmogorov, 
		C.~R.~Acad. Sci. USSR {\bf 30}, 301; 538 (1941).
\bibitem{Benzi84} R.~Benzi, G.~Paladin, G.~Parisi and A.~Vulpiani, 
		J. Phys. A: Math. Gen. {\bf 17}, 3521 (1984).
\bibitem{Frisch78} U.~Frisch, P-L.~Sulem and M.~Nelkin,
		J. Fluid Mech. {\bf 87}, 719 (1978).
\bibitem{Lyra98} M.~L.~Lyra and C.~Tsallis, 
		Phys. Rev. Lett. {\bf 80}, 53 (1998).
\bibitem{Costa} U.M.S.~Costa, M.L.~Lyra, A.R.~Plastino and
		C.~Tsallis, Phys. Rev. E {\bf 56}, 245 (1997).
\bibitem{Chhabra91} A.B.~Chhabra and K.R.~Sreenivasan, 
		Phys. Rev. A {\bf 43}, 1114 (1991).
\bibitem{Havrda-Charvat} J.~H.~Havrda and F.~Charvat,
		Kybernatica {\bf 3}, 30 (1967).
\bibitem{Beck00} C.~Beck, Physica A {\bf 277}, 115 (2000).
\bibitem{Beck-Lewis-Swinney01} C.~Beck, G.S.~Lewis and H.L.~Swinney, Phys. Rev. 
		E {\bf 63}, 035303-1 (2001).
\bibitem{Beck01preprint} C.~Beck (2001) preprint.
\bibitem{Fujisaka} T.~Watanabe, Y.~Nakayama and H.~Fujisaka, Phys. Rev. E {\bf 61},
		R1024 (2000).


\end{thebibliography}
\end{document}